\documentstyle[12pt,prl,aps,multicol,psfig,epsf]{revtex}
\begin{document}

\baselineskip=7mm
\def\ap#1#2#3{           {\it Ann. Phys. (NY) }{\bf #1} (19#2) #3}
\def\arnps#1#2#3{        {\it Ann. Rev. Nucl. Part. Sci. }{\bf #1} (19#2) #3}
\def\cnpp#1#2#3{        {\it Comm. Nucl. Part. Phys. }{\bf #1} (19#2) #3}
\def\apj#1#2#3{          {\it Astrophys. J. }{\bf #1} (19#2) #3}
\def\asr#1#2#3{          {\it Astrophys. Space Rev. }{\bf #1} (19#2) #3}
\def\ass#1#2#3{          {\it Astrophys. Space Sci. }{\bf #1} (19#2) #3}

\def\apjl#1#2#3{         {\it Astrophys. J. Lett. }{\bf #1} (19#2) #3}
\def\ass#1#2#3{          {\it Astrophys. Space Sci. }{\bf #1} (19#2) #3}
\def\jel#1#2#3{         {\it Journal Europhys. Lett. }{\bf #1} (19#2) #3}

\def\ib#1#2#3{           {\it ibid. }{\bf #1} (19#2) #3}
\def\nat#1#2#3{          {\it Nature }{\bf #1} (19#2) #3}
\def\nps#1#2#3{          {\it Nucl. Phys. B (Proc. Suppl.) } {\bf #1} (19#2) #3}
\def\np#1#2#3{           {\it Nucl. Phys. }{\bf #1} (19#2) #3}

\def\pl#1#2#3{           {\it Phys. Lett. }{\bf #1} (19#2) #3}
\def\pr#1#2#3{           {\it Phys. Rev. }{\bf #1} (19#2) #3}
\def\prep#1#2#3{         {\it Phys. Rep. }{\bf #1} (19#2) #3}
\def\prl#1#2#3{          {\it Phys. Rev. Lett. }{\bf #1} (19#2) #3}
\def\pw#1#2#3{          {\it Particle World }{\bf #1} (19#2) #3}
\def\ptp#1#2#3{          {\it Prog. Theor. Phys. }{\bf #1} (19#2) #3}
\def\jppnp#1#2#3{         {\it J. Prog. Part. Nucl. Phys. }{\bf #1} (19#2) #3}

\def\rpp#1#2#3{         {\it Rep. on Prog. in Phys. }{\bf #1} (19#2) #3}
\def\ptps#1#2#3{         {\it Prog. Theor. Phys. Suppl. }{\bf #1} (19#2) #3}
\def\rmp#1#2#3{          {\it Rev. Mod. Phys. }{\bf #1} (19#2) #3}
\def\zp#1#2#3{           {\it Zeit. fur Physik }{\bf #1} (19#2) #3}
\def\fp#1#2#3{           {\it Fortschr. Phys. }{\bf #1} (19#2) #3}
\def\Zp#1#2#3{           {\it Z. Physik }{\bf #1} (19#2) #3}
\def\Sci#1#2#3{          {\it Science }{\bf #1} (19#2) #3}

\def\n.c.#1#2#3{         {\it Nuovo Cim. }{\bf #1} (19#2) #3}
\def\r.n.c.#1#2#3{       {\it Riv. del Nuovo Cim. }{\bf #1} (19#2) #3}
\def\sjnp#1#2#3{         {\it Sov. J. Nucl. Phys. }{\bf #1} (19#2) #3}
\def\yf#1#2#3{           {\it Yad. Fiz. }{\bf #1} (19#2) #3}
\def\zetf#1#2#3{         {\it Z. Eksp. Teor. Fiz. }{\bf #1} (19#2) #3}
\def\zetfpr#1#2#3{         {\it Z. Eksp. Teor. Fiz. Pisma. Red. }{\bf #1} (19#2) #3}
\def\jetp#1#2#3{         {\it JETP }{\bf #1} (19#2) #3}
\def\mpl#1#2#3{          {\it Mod. Phys. Lett. }{\bf #1} (19#2) #3}
\def\ufn#1#2#3{          {\it Usp. Fiz. Naut. }{\bf #1} (19#2) #3}
\def\sp#1#2#3{           {\it Sov. Phys.-Usp.}{\bf #1} (19#2) #3}
\def\ppnp#1#2#3{           {\it Prog. Part. Nucl. Phys. }{\bf #1} (19#2) #3}
\def\cnpp#1#2#3{           {\it Comm. Nucl. Part. Phys. }{\bf #1} (19#2) #3}
\def\ijmp#1#2#3{           {\it Int. J. Mod. Phys. }{\bf #1} (19#2) #3}
\def\ic#1#2#3{           {\it Investigaci\'on y Ciencia }{\bf #1} (19#2) #3}
\def\tp{these proceedings}
\def\pc{private communication}
\def\ip{in preparation}
\relax
\newcommand{\TeV}{\,{\rm TeV}}
\newcommand{\GeV}{\,{\rm GeV}}
\newcommand{\MeV}{\,{\rm MeV}}
\newcommand{\keV}{\,{\rm keV}}
\newcommand{\eV}{\,{\rm eV}}
\newcommand{\Tr}{{\rm Tr}\!}
\renewcommand{\arraystretch}{1.2}
\newcommand{\beq}{\begin{equation}}
\newcommand{\eeq}{\end{equation}}
\newcommand{\beqa}{\begin{eqnarray}}
\newcommand{\eeqa}{\end{eqnarray}}
\newcommand{\ba}{\begin{array}}
\newcommand{\ea}{\end{array}}
\newcommand{\bmat}{\left(\ba}
\newcommand{\emat}{\ea\right)}
\newcommand{\refs}[1]{(\ref{#1})}
\newcommand{\ler}{\stackrel{\scriptstyle <}{\scriptstyle\sim}}
\newcommand{\ger}{\stackrel{\scriptstyle >}{\scriptstyle\sim}}
\newcommand{\lag}{\langle}
\newcommand{\rag}{\rangle}
\newcommand{\ns}{\normalsize}
\newcommand{\cm}{{\cal M}}
\newcommand{\gr}{m_{3/2}}
\newcommand{\p}{\partial}
\def\321{$SU(3)\times SU(2)\times U(1)$}
\def\emt{$L_e+L_\mu-L_\tau-L_s$~}
\def\ls{$U(1)_S$}
\def\tl{{\tilde{l}}}
\def\tL{{\tilde{L}}}
\def\bd{{\overline{d}}}
\def\tL{{\tilde{L}}}
\def\a{\alpha}
\def\b{\beta}
\def\g{\gamma}
\def\c{\chi}
\def\d{\delta}
\def\D{\Delta}
\def\db{{\overline{\delta}}}
\def\Db{{\overline{\Delta}}}
\def\e{\epsilon}
\def\l{\lambda}
\def\n{\nu}
\def\m{\mu}
\def\nt{{\tilde{\nu}}}
\def\p{\phi}
\def\P{\Phi}
\def\k{\kappa}
\def\x{\xi}
\def\r{\rho}
\def\s{\sigma}
\def\t{\tau}
\def\th{\theta}
\def\ne{\nu_e}
\def\nm{\nu_{\mu}}
\def\snui{\tilde{\nu_i}}
\def\la{{\makebox{\tiny{\bf loop}}}}
\renewcommand{\Huge}{\Large}
\renewcommand{\LARGE}{\Large}
\renewcommand{\Large}{\large}
\title{
\hfill hep-ph/0110272\\[1.5 cm] {\bf Neutrino Anomalies and}
\\{\bf Quasi-Dirac Neutrinos}}
\author{ Srubabati Goswami~$^{a}$ and Anjan S. Joshipura~$^{b}$\\
{\ns\it $(b)$~ Theoretical Physics Group, Physical Research Laboratory,}\\
{\ns\it Navarangpura, Ahmedabad, 380 009, India.}\\
{\ns\it $(a)$~ Saha Inst. of Nuclear Physics}\\
{\ns\it 1/AF, Bidhannagar,Calcutta,700 064, India}}
 \maketitle
\begin{abstract}

We discuss possibility of describing solar, atmospheric and LSND
results with four neutrinos forming two quasi-degenerate pairs.
The simplest versions of this $2+2$ scheme with either $\nu_e$ or
$\nu_\mu$ mixing exclusively with sterile neutrino is disfavored
by the SNO and atmospheric neutrino results respectively. A
generalized scheme with sterile state participating in both the
solar and atmospheric oscillations is still allowed. We show that
the complex pattern of mixing needed for this purpose follows from
a simple \emt symmetry. Specific form of \emt symmetric mass
matrix is determined from experimental results. Two theoretical
schemes which lead to this form and a  proper breaking of \emt
symmetry are discussed.
\end{abstract}

\section{Introduction}

Both positive and negative results on neutrino oscillation
searches have provided very important clues on possible patterns
of neutrino masses and mixing \cite{rev}. These results include
experiments detecting solar and atmospheric neutrinos as well as
laboratory experiments such as LSND, CHOOZ etc. The observed solar
and atmospheric neutrino deficits provide concrete ground to
believe in neutrino oscillations. Along with the LSND results,
they  give important and by now well-known \cite{rev} information
on neutrino masses.

The latest input in the analysis of neutrino spectrum is results
of solar neutrino deficit seen at SNO \cite{sno}. This experiment
finds lower neutrino flux in their charged current events compared
to the flux inferred from the elastic scattering at SuperKamioka
\cite{sk1} which receives contributions from the charged as well
as the neutral current processes. The difference in these two
fluxes is consistent with $\nu_e$ predominantly converting to
active flavours only. The global analysis of the solar neutrino
data \cite{global} shows that complete conversion of the solar
$\nu_e$ to sterile neutrino is a disfavored possibility allowed at
$3\sigma$ level that too  mainly in case of the vacuum
oscillations only.

The implications of the above results for the neutrino masses are as
follows.
\begin{itemize}

\item The oscillation interpretation of data requires \cite{rev}
three different mass scales ($\Delta_{LSND}\sim \eV^2$,
$\Delta_{A}\sim (5-8)\cdot 10^{-3} \eV^2$ and $\Delta_{S}\leq
10^{-4}\eV^2$) and four light neutrinos to account for these
(mass)$^2$ differences.
\item The schemes with hierarchical masses is highly disfavored
from the point of view of explaining all neutrino anomalies
\cite{grimus,31}. In this scheme, the allowed LSND probability is
found to be smaller than the observed \cite{lsnd} one when
negative results of neutrino oscillation searches at Bugey and
CDHS are taken into consideration.
\item All the experimental results before  SNO could be understood
\cite{grimus} in terms of a simple picture in which four neutrinos
group themselves into two pairs with
a gap of order $\sqrt{\Delta_{LSND}}$. There are two versions:\\
\noindent (a) in which one of the pairs consists of $\nu_e,\nu_s$
and accounts for the solar neutrino anomaly. The other is
responsible
for $\nu_\m-\nu_\tau$ oscillations of atmospheric neutrinos.\\
\noindent (b) corresponds to converse possibility with
$\nu_\m-\nu_s$ accounting for the atmospheric neutrino deficit,
solar neutrinos converting themselves  completely to active
components. Both these possibilities have been termed as $2+2$
schemes\cite{grimus,sp} of neutrino masses.
\end{itemize}

The above possibilities can be incorporated naturally into
theoretical schemes for neutrino masses and there have been number
of models \cite{models,rsym,singlet,maroy,gaur} realizing these
possibilities. The possibility (b) was already disfavored by the
observed absence \cite{sk2} of the matter effects in atmospheric
neutrino data. Now the SNO results \cite{global} strongly
disfavors (a). Thus it becomes a challenging task to understand
experimental results within the four neutrino scenario and build
necessary framework to account for this. We discuss simple four
neutrino schemes in this paper which can simultaneously explain
the solar, atmospheric and LSND results.

While the possibilities (a) and (b) above are the simplest
realization of the 2+2 schemes they are not exhaustive. In
general, a sterile state can simultaneously but partially
influence both the solar and atmospheric neutrino oscillations.
The most general possibility in this context was worked out in
\cite{sp}. It was shown that the LSND results can be accounted for
in 2+2 scheme without conflicting with the negative searches of
neutrino oscillations if $\nu_e$ and $\nu_\m$ reside mainly in
different mass pairs. The unitarity of mixing matrix then dictates
the following mixing pattern among four neutrinos \cite{sp}:

\beqa \label{gen} \nu_e&=& \cos\th_S ~\nu_1+\sin\th_S ~\nu_2
+{\cal O}(\e) ~,\nonumber \\
\nu_\m&=&\cos\th_A ~\nu_3+\sin\th_A ~\nu_4+{\cal O}(\e)~, \nonumber \\
\nu_\t&=& \sin \a (-\sin\th_A ~\nu_3+\cos\th_A ~\nu_4)+
          \cos\a (-\sin\th_S ~\nu_1+\cos\th_S ~\nu_2)+{\cal O}(\e)~, 
\nonumber \\
\nu_s&=& \cos\a (-\sin\th_A ~\nu_3+\cos\th_A ~\nu_4)-
          \sin\a (-\sin\th_S ~\nu_1+\cos\th_S ~\nu_2)+{\cal
O}(\e)~.  \eeqa

The  $\nu_{1,2}$ and $\nu_{3,4}$ represent two quasi-Dirac pairs
of neutrinos with definite masses $m_{\nu_i}~(i=1,4)$.
$\theta_S,\theta_A$ respectively denote mixing angles relevant for
the solar and atmospheric neutrino oscillations. The angle $\a$
determines \footnote{$\a$ defined here differs from the one in
\cite{sp} by $\pi/2$.} the amount of sterile component in these
two fluxes. The splittings $\Delta_{12},\Delta_{34}$
($\Delta_{ij}\equiv m_{\nu_j}^2-m_{\nu_i}^2$) respectively account
for the solar and atmospheric oscillations. As is seen, the
$\nu_e$ and $\nu_\m$ reside in different mass pairs apart from
small ${\cal O}(\e) $ corrections. The parameter $\epsilon$ thus
decides the amplitude for the LSND oscillations.

Two examples of pure active sterile mixing discussed above
correspond to special cases of eq.(\ref{gen}) with $\a=\pi/2$
(case (a)) and $ \a=0$ (case (b)). While these extreme cases are
ruled out, intermediate possibility with non-zero $\a$ is still
allowed. The phenomenology of this case was studied in
\cite{sp,gar1} and was updated after SNO results in \cite{gar2}.
Basically, the same parameter $\a$ determines amount of sterile
component in the solar as well as atmospheric neutrino flux and
thus can be constrained by both experiments. It was found in
\cite{gar2} that the  solar as well as atmospheric data can be
fitted with a non-zero $\a$ in two possible ways. Either $\a$ is
large in which case the sterile state mainly appears in the solar
neutrino flux. The best fit in this case is obtained for

\beq \sin^2\a\approx 0.8 ~.\eeq The possibility (a) above is
allowed in this case only at 99\%CL. The other case corresponds to
small $\a$ and sterile component mainly in the atmospheric
neutrino flux. The best fit value of $\a$ is given in this case by

\beq \sin^2\a\approx 0.1 ~.\eeq

The possibility (b) above can be allowed around 95\%CL in this
case. It follows that complete conversion of $\nu_e$ to $\nu_s$ in
the solar flux or $\nu_\m$ to $\nu_s$ in the atmospheric flux is a
disfavored possibility although best fit value of $\a$ is not very
far from these two limiting cases. General mixing pattern in
eq.(\ref{gen}) is a more favored possibility. A priori, this
mixing pattern looks complex but as we discuss below, it follows
if $4X4$ neutrino mass matrix respects a \emt symmetry. We write
down a specific ansatz based on this symmetry in the next section
where we also show that \emt symmetric mass matrix can be
determined from the experimental results. Next two sections are
devoted to realization of the mass matrix based on \emt symmetry.
Last section summarizes the results obtained.

\section{\emt symmetry and neutrino mixing}
Let us parameterize the most general $4X4$ matrix in the basis
$(\nu_e,\nu_\m,\nu_\tau,\nu_s)$ as follows:
 \beq {\cal M}_\nu=\left ( \ba{cc} \Delta&{\mathbf m}\\
         {\mathbf m}^T&\Delta' \\ \ea \right)~. \label{ansatz}\eeq
Here each sub-block is a $2X2$ matrix. The existence of
pseudo-Dirac pairs implies some partially broken $U(1)$ symmetry.
Such symmetry follows if $2X2$ matrices $\Delta,\Delta'$ are
sub-dominant compared to ${\mathbf m}$ in eq.(\ref{ansatz}). Let
us choose $\Delta=\Delta'=0$ as a first approximation. Then for
arbitrary ${\mathbf m}$, ${\cal M}_\nu$ is invariant under
$L_e+L_\m-L_s-L_\tau$ symmetry. It is possible to exactly
diagonalize eq.(\ref{ansatz}) in this case. This is done in two
steps. First, ${\mathbf m}$ is  diagonalized by the following
$2X2$ rotations R:

 \beq
R_\beta~{\mathbf m}~R^T_\alpha =Diag. (m_1,m_2)~, \label{2by2}\eeq
where $\a,\b$ denote the angles of rotations and $m_{1,2}$ are the
eigenvalues of ${\mathbf m}$. Given eq.(\ref{2by2}), the following
$4X4$ matrix diagonalizes ${\cal M}_\nu$ when $\Delta,\Delta'$ are
zero:
\beq R' R{\cal M}_\nu R^T R'^T=Diag. (m_{1},-m_{1},m_{2},-m_{2})~,
\eeq
where,
\beq \label{u} R=\bmat{cc}
R_\beta&0\\
0&R_\alpha\\ \emat ~,\eeq and

 \beq \label{uprime}
 R'=\left( \ba{cccc}
 {1\over \sqrt{2}}&0& {1\over \sqrt{2}}&0\\
{1\over \sqrt{2}}& 0& -{1\over \sqrt{2}}&0\\
0& {1\over \sqrt{2}}&0&{1\over \sqrt{2}}\\
0&{1 \over \sqrt{2}}&0&-{1\over \sqrt{2}}\\
\ea \right)~.\eeq

Eqs.(\ref{u},\ref{uprime}) together imply the following mixing
among neutrinos:
\beq \nu_\a \equiv U_{\a i}\nu_i=(R'~R)^T_{\a i}~\nu_i ~. \label{udef}\eeq
Explicitly,
 \beqa \label{mixing}
 \nu_e&=&{c_\b\over\sqrt{2}} (\nu_1+\nu_2)
+{s_\b \over \sqrt{2}} (\nu_3+\nu_4)~, \nonumber \\
\nu_\m&=& -{s_\b\over \sqrt{2}} (\nu_1+\nu_2)
+{c_\b \over \sqrt{2}} (\nu_3+\nu_4)~,\nonumber \\
\nu_\tau&=&{c_\a\over\sqrt{2}} (\nu_1-\nu_2)
+{s_\a \over \sqrt{2}} (\nu_3-\nu_4)~, \nonumber \\
\nu_s&=&-{s_\a\over\sqrt{2}} (\nu_1-\nu_2) +{c_\a \over \sqrt{2}}
(\nu_3-\nu_4) ~.\eeqa

The above pattern coincides with  eq.(\ref{gen}) if $s_\b\ll 1$
and $\theta_A=\theta_S=\frac{\pi}{4}$. It is easily seen that $\b$
governs the amplitude of LSND oscillations. Eq.(\ref{mixing})
implies
$$ \sin^2 2 \theta_{LSND}\approx 4 (U_{e4} U_{\m4}+U_{e3}U_{\m 3})^2
=\sin^2 2\beta$$
when $m_{\nu_{1,2}}\ll m_{\nu_{3,4}}\sim \eV$. The observations at
LSND then implies  that $\sin^2 2\b\sim 3\cdot 10^{-3}$.
$\theta_A=\theta_S$=$\pi/4$ is a prediction of the model which
arise as a consequence of the \emt symmetry. The perturbation
$\Delta,\Delta'$ which would cause the splitting of the neutrino
masses would also change this prediction somewhat but one will get
two large mixing angles needed on phenomenological grounds.

Eq.(\ref{mixing}) leads to the following survival probability in
disappearance experiment such as CHOOZ \cite{chooz}

\beqa P_{ee}&=&1-4 U_{e3}^3 U_{e4}^2 \sin^2 {\Delta_{34}t \over
4E}-2 (U_{e1}^2+U_{e2}^2)(U_{e3}^2+U_{e4}^2)
\nonumber \\
&\approx&1-s_{\b}^4{\Delta_{34}t \over 4E}-\frac{1}{2}\sin^2
2\beta ~.\eeqa

This probability is correlated with the LSND angle. Its
oscillatory part is suppressed in view of the LSND results on
$s_\b$. The average term is smaller than the present limit set by
CHOOZ \cite{chooz} experiment but it can be significant.

It is clear that the \emt symmetric mass matrix leads to
phenomenologically desirable pattern naturally. Its breaking is
needed to generate the mass splittings. Details depend upon the
form and strength of the perturbation $\Delta,\Delta'$ which is
yet unspecified. But the following observation is relevant. One
can calculate the (mass)$^2$ differences using perturbation theory
assuming that $\Delta,\Delta'$ have general structure with all
elements having equal strength. Using the unperturbed
eigenfunctions following from $U$ in eq.(\ref{udef}), we get
\beqa \Delta_{A} &\equiv& (m_{\nu_4}^2-m_{\nu_3}^2)\approx m_2
{\cal O} (\delta)
\nonumber \\
\Delta_{S} &\equiv& (m_{\nu_2}^2-m_{\nu_1}^2)\approx m_1 {\cal O}
(\delta) ~.\eeqa

It follows that
 \beq \label{ratio} {\Delta_{S}\over
\Delta_{A}}\approx{m_1\over m_2} {\cal O}(1) ~.\eeq
Thus, hierarchy in the solar and atmospheric scales is linked in
this case to intra-splitting between two pairs and one generically
needs $m_1\ll m_2$. We shall present an example where
eq.(\ref{ratio}) is realized naturally with ${\cal O}(1)$
parameter being exactly 1.
${\cal M}_\nu$ is characterized by four parameters $\a,\b,m_1$ and
$m_2$. All these parameters are approximately determined
phenomenologically: $m_2$ and $\b$ from LSND results, $\a$ from
general fit \cite{gar2}  to solar and atmospheric data and $m_1$
from eq.(\ref{ratio}). Thus ${\cal M}_\nu$ gets phenomenologically
determined in the symmetric limit.

Let us parameterize ${\mathbf m}$ as follows: \beq {\mathbf
m}=\left (
\ba{cc} a_1&a_2\\
         A_1&A_2 \\ \ea \right)~. \label{md}\eeq

The elements $a_i$ and $A_i$ ($i=1,2$) are determined in terms of
mixing and masses as follows:

\beqa \label{elements} A_2&=&m_2 c_\b c_\a+m_1 s_\b s_\a\approx
m_2 c_\a ~, \nonumber \\ A_1&=&m_2 s_\a c_\b-m_1 s_\b c_\a\approx
m_2 s_\a ~, \nonumber \\ a_2&=&m_2 c_\a s_\b-m_1 s_\a c_\b ~,\nonumber
\\ a_1&=&m_2 s_\b s_\a+m_1 c_\b c_\a ~.\eeqa

The parameters on the RHS are directly determined from
experiments. Since both $s_\b$ and $m_1$ are required to be small,
the above equations imply that
 \beq \label{rel1}
\frac{a_1}{A_1}\sim\frac{a_2}{A_2}\ll 1 \eeq
and also
\beq \label{rel2} {a_1\over A_1}-{a_2\over A_2}\approx {\cal
O}(1)~{\Delta_S\over s_\a c_\a \Delta_A} .\eeq which itself is
small. Successful model should realize the above hierarchy and we
present two specific examples.
\section{Radiative scheme}
We showed that \emt symmetry leads to phenomenologically
consistent $2+2$ model and identified a specific structure for the
four neutrino mass matrix in this limit. We now discuss a
radiative scheme which realizes this structure and also provides
necessary breaking of the \emt symmetry. Our starting point is the
observation that a part of \emt symmetry namely,
$\L_e+\L_\m-L_\tau$ already arise as an accidental and approximate
symmetry in the Zee model \cite{zee} of neutrino masses. Thus it
is natural to start with this model. We add a singlet neutrino
$\nu_s$ to it and impose an associated singlet lepton number
symmetry $U(1)_S$ which is carried by $\nu_s$, rest of the
fermions remaining unchanged under it. Due to inherent
$L_\e+\L_\m-L_\tau$ symmetry of the Zee model, the radiatively
generated $4X4$ matrix automatically displays approximate \emt
symmetry although we start with only a $U(1)_S$ symmetry. The
Higgs fields of the model are two doublets $\phi_{1,2}$ and a
charged singlet $h_1^+$ as in Zee model \cite{zee} and an
additional charged singlet $h_2^+$ carrying the same $U(1)_S$
charge as $\nu_s$. The leptonic Yukawa couplings in the model are
then given by

\beq \label{yukawa} - {\cal
L}_Y=f_{ij}\bar{{\ell}}_{iL}^c~{\ell}_{jL}~h_1^+
+\b_i\bar{\nu_s}e_{iR}~h_2^+ +g_{ij}\bar{\ell
}_{iL}~e_{jR}\tilde{\phi_1}+H.c.  ~.\eeq
Note that the $U(1)_S$ symmetry forbids the Dirac coupling between
${\ell}_L$ and $\nu_s$ as well as the Majorana mass for the
$\nu_s$. Above Yukawa couplings are automatically invariant under
Lepton number as in Zee model. Lepton number as well as $U(1)_S$
symmetry is broken softly in Higgs sector through the following
terms:
\beq \label{soft} \m^2 h_1^+h_2^- +\gamma_i\phi_1\phi_2 h_i^+ ~.\eeq
These soft symmetry breaking terms radiatively generate neutrino
masses. The mass matrix generated at the 1-loop level has the
following structure:

\beq \label{zeem}{\cal{M}_\nu} =\bmat{cccc}
0&\d&a_1&a_2\\
\d&0&A_1&A_2\\
a_1&A_1&0&\d'\\
a_2&A_2&\d'&0 \\ \emat~, \eeq

where

\beqa \label{aizee} a_1\sim {C\over v_1} f_{e\tau} m_{\tau}^2
&~~;~~& A_1\sim {C\over v_1}
 f_{\mu\tau} (m_{\tau}^2-m_\mu^2)~, \nonumber \\
a_2\sim C' (f_{e\tau} \b_\tau m_{\tau}+f_{e\mu}\b_\m m_\m)&~~;~~&
A_2\sim C' f_{\m\tau}\b_\tau m_{\tau}~, \nonumber \\
\d\sim  {C\over v_1} f_{e\m} m_{\m}^2 &~~;~~& \d'\sim C'
f_{\tau\m}\b_\m m_{\m}~, \eeqa
where we have neglected the electron mass. $C$ and $C'$ in the
above equations are given by
 \beq \ba{cc}
 C\sim {1\over 16 \pi^2}
f(m_h)~~~~&;~~~~C'\sim {1\over 16 \pi^2} g(m_h)\\ \ea \eeq
$f$ and $g$ are dimensionless functions of parameters in Higgs
potential (denoted collectively by $m_h$) including the soft
symmetry breaking terms displayed in eq.(\ref{soft}).

Note that the $\d,\d'$ are determined by the muon mass and rest by
the tau as well as muon mass. Thus, eq.(\ref{zeem}) has the
perturbative structure required in our ansatz, eq.(\ref{ansatz}).
One can calculate splitting among neutrino masses generated by
${\cal M}_\nu$ in eq.(\ref{zeem}). Treating $\d,\d'$ as
perturbation and using the unperturbed eigenfunctions given in
eq.(\ref{mixing}), we get
\beqa m_{\nu_1}&\sim& m_1-(\d c_\b s_\b+\d'c_\a s_\a)~, \nonumber \\
m_{\nu_2}&\sim&-m_1-(\d c_\b s_\b+\d'c_\a s_\a) ~, \nonumber \\
m_{\nu_3}&\sim& m_2+(\d c_\b s_\b+\d'c_\a s_\a)~,\nonumber \\
m_{\nu_4}&\sim&-m_2+(\d c_\b s_\b+\d'c_\a s_\a) ~. \eeqa

This leads to the following splittings:

\beqa \label{zeesplitting} \Delta_A&\approx& -4 m_2 (\delta c_\b
s_\b+\d'c_\a s_\a)~,
\nonumber \\
{\Delta_S\over \Delta_A}&\approx&-{m_1\over m_2}~. \eeqa

The hierarchy in the solar and atmospheric scale is insensitive to
strength of perturbation and is solely determined by the ratio of
masses of the Dirac pairs. Eq.(\ref{aizee}) leads to

$$ \frac{a_1}{A_1}\approx \frac{a_2}{A_2}\approx
\frac{f_{e\tau}}{f_{\m \tau}}~. $$

The required hierarchy in eq.(\ref{rel1}) can be obtained if
$f_{e\tau}$ and $f_{\mu\tau}$ are hierarchical. Eq.(\ref{rel2})
now assumes the following form
\beq ({a_1\over A_1}-{a_2\over A_2})\sim {f_{e\tau}\over
f_{\m\tau}}{m_\mu^2\over m_{\tau}^2}-{\b_\m\over
\b_\tau}{f_{e\m}\over f_{\m\tau}}{m_\m\over m_\tau}~.\eeq
The difference on the LHS can thus be naturally small. It is
possible to determine the basic parameters
$f_{e\mu},f_{e\tau},f_{\mu\tau}$ and $\b_{\mu,\tau}$ diretcly
using eqs.(\ref{elements},\ref{zeesplitting}). Little algebra
gives,
\beqa \label{couplings}
\b_\tau&\approx&\frac{C}{C'}\frac{m_\tau}{v_1}\frac{c_\a}{s_\a}~~~;~~~
\frac{\b_\mu}{\b_{\tau}}\approx
-\frac{\Delta_A}{m_2^2}\frac{m_\tau}{m_\mu}\frac{1}{s_\a c_\a^2}
~,\nonumber \\
f_{\mu\tau}&\approx&\frac{m_2v_1}{m_\tau^2}\frac{s_\a}{C}~~~;~~~
\frac{f_{e\tau}}{f_{\mu\tau}}\approx s_\b
+\frac{\Delta_S}{\Delta_A}\frac{c_\a}{s_\a}
~,\nonumber \\
~~~~~~~~~~~~~~~~~~~~~~~~\frac{f_{e\mu}}{f_{\mu\tau}}&\approx&\frac{\Delta_S}{\Delta_A}\frac{m_2^2}{\Delta_A}
c_\a ~.\eeqa

Eqs.(\ref{elements},\ref{zeesplitting}) reveal that the required
magnitudes of $a_{1,2}$ may be comparable to the perturbation
$\d,\d'$. In this case, some of the perturbative  results may
change. It is therefore appropriate to perform exact
diagonalization of matrix in  eq.(\ref{zeem}). We have done this
numerically
choosing values of parameters  around the ones given in
eq.(\ref{couplings}). The specific values chosen are \beq
\label{values} \ba{ccc}
\b_\mu=-4.3 \cdot 10^{-3};&~~~~~~~~~~~~~&\b_{\tau}= 2.7 \cdot 10^{-2}\\
 f_{e\mu}=2.1 \cdot
10^{-5};&~~~~f_{e\tau}=3.6 \cdot 10^{-7};~~~~~&f_{\mu\tau}=
8.6 \cdot 10^{-6} \\

\ea~. \eeq
We chose  $C\sim C'\sim 0.01$ for definiteness. The above choice
when substituted in eq.(\ref{zeem}) leads to the following values
for the solar and atmospheric scales: \beqa
\Delta_S&=&2.3 \cdot 10^{-5}\eV^2 ~,\nonumber \\
\Delta_A&=&3.1 \cdot 10^{-3} \eV^2 ~,\nonumber \\
\Delta_{LSND}&=&0.3 \eV^2~.\eeqa

The mixing matrix is given by \beq U^T\approx \bmat{cccc}
0.826&0.57&0.02&0.017\\
-0.024&-0.011&0.71&0.71\\
0.52&0-.75&0.24&-0.28\\
-0.23&0.33&.64&-0.65\\ \emat~. \eeq

This can be seen to correspond to the following values for various
parameters:

\beq \ba{ccc}
\sin^2 2\theta_A=0.99;&~~&\sin^2 2\theta_S=0.88  \\
\sin^2 2\theta_{LSND}=2.7\cdot 10^{-3};&~~& \sin^2 2
\theta_{chooz}=1.4\cdot 10^{-3} \\
~~~~~~~~~~~~~~~~~~~~~~~~~~~~~~~~~~~~~~~~~~~~~~~~~s_\a^2\approx 0.15 &~&\ea
\eeq
The solar mixing angle is reduced significantly compared to its
maximal value in the absence of perturbation. This is welcome
since strictly maximal mixing is not favored at least in the two
generation analysis of the solar data \cite{global}. Clearly,
there would be ranges in the basic parameters of the model which
would reproduces the correct mixing and masses.

We note that the structure of the mass matrix in eq.(\ref{zeem})
coincides with the one discussed in \cite{gaur} but the underlying
model presented here is much simpler. More importantly,
phenomenological emphasis here is very different. We have shown
that the basic structure in eq.(\ref{zeem}) displaying approximate
\emt symmetry provides a concrete realization of the generalized
$2+2$ model which is fully consistent with all the neutrino
anomalies even after inclusion of SNO results.
\section{Seesaw model}
We now discuss how the ansatz of section (2) can be derived in
seesaw type scheme. There are two ways of obtaining a light
sterile neutrino in seesaw model. One is to assume that the mass
matrix of the right handed neutrino is singular \cite{glashow}.
This can be done through some symmetry \cite{asjr}. The massless
RH neutrino resulting from this singular matrix picks up a mass
through its Dirac coupling with the active neutrino. The RH
neutrino remains strictly massless and can provide the sterile
state if its Dirac coupling is also forbidden by a symmetry.
Example of this was recently presented in \cite{mohapatra}.
Alternative possibility is to add a sterile state to the
conventional seesaw picture and impose symmetry which keeps it
light. Examples of this possibility were discussed in
\cite{rsym,maroy}. Consider an active state $\nu_{L}$, its RH
partner $\nu_{R}$ and a sterile (left-handed) state $\nu_s$ with
the following mass matrix in the basis
$(\nu_L,\nu_s,\nu_R^c)$:

\beq \label{light} \bmat{ccc} 0&0&m_D\\
                0&0&m_S\\
                m_D&m_S&M\\ \emat~ \eeq

This mass matrix leads to a massless, a light ($\sim
\frac{m_D^2+m_S^2}{M}$) and a heavy ($\sim M$) neutrino. The
sterile state mixes with the active state and influences the
phenomenology. Crucial points to note are the absence of Dirac
mass term between $\nu_s$ and $\nu_L$ and the absence of the
Majorana mass term for $\nu_s$. This can be achieved by means of
some symmetry, e.g. $R$ symmetry as in \cite{rsym}. This matrix
was proposed as a model for solving the solar neutrino anomaly
through $\nu_e-\nu_s$ mixing. Now we generalize the above idea to
obtain the ansatz discussed in the last section.

We consider the conventional seesaw picture with three left-handed
and three right-handed neutrinos and add to it a sterile state
$\nu_s$ which would remain light. The lightness can be ensured by
a structure which is generalization of eq.(\ref{light}) to three
generations. We demand separate conservation of $L_e+L_\m-L_\tau$
and the lepton number $L_s$ corresponding to sterile state
$\nu_s$. This would lead to \emt symmetric $4X4$ matrix of the
ansatz, eq.(\ref{ansatz}). We however need to break this symmetry
softly in order to obtain realistic mass spectrum. Our model thus
has two sets of mass terms: ${\cal L}_m$ which respect the
symmetry and ${\cal L'}_m$ which break it softly. The symmetric
part is given by
\beqa \label{sym} -{\cal L}_m&=&m_{ee}\bar{\nu}_{eL}\nu_{eR}+
m_{e\m}\bar{\nu}_{eL}\nu_{\mu R}+m_{\mu e} \bar{\nu}_{\mu L}
\nu_{eR}\nonumber \\
&+& m_{\mu\mu}\bar{\nu}_{\mu L}\nu_{\mu R}+
m_{\tau\tau}\bar{\nu}_{\tau L}\nu_{\tau R}+~~~~ H.c. \nonumber \\
&+&\frac{1}{2} (M_{e\tau} \bar{\nu}_{eR}^c\nu_{\tau R}+M_{e\tau}
\bar{\nu}_{\tau R}^c\nu_{eR}+M_{\mu\tau} \bar{\nu}_{\mu
R}^c\nu_{\tau R}+M_{\mu\tau} \bar{\nu}_{\tau R}^c\nu_{\mu R}) ~.
\eeqa
Due to the combined effects of two $U(1)$ symmetries, the $\nu_s$
remains massless and decoupled from rest of the fermions in
eq.(\ref{sym}). Its couplings  with active neutrinos are induced
entirely by the soft symmetry breaking sector ${\cal L'}_m$ chosen
as follows:
\beqa \label{break} -{\cal L'}_m&=&p_e\bar{\nu}_s\nu_{eR}+p_\mu
\bar{\nu}_s\nu_{\mu R}+p_\tau \bar{\nu}_s \nu_{\tau R}~~~+H.c.
\nonumber \\
&+&\frac{1}{2}M_{ee} \bar{\nu}_{eR}^c\nu_{eR} ~. \eeqa
Note that all terms in the above equation connect only the sterile
states and are soft in the technical sense- they would not lead to
any divergences \footnote{ Models using such soft symmetry
breaking were proposed in \cite{gl}.}. Arbitrary choice of soft
terms is consistent with this technical requirement. We have added
all possible singlet mass terms above except for the direct mass
term $\bar{\nu}_s^c\nu_s$ and some additional mass terms among the
RH neutrinos. Addition of the latter terms do not make any
qualitative change. The omission of $\bar{\nu}_s^c\nu_s$ can be
justified if soft terms are assumed to come through normalizable
couplings of sterile state with standard model singlets
$\eta_1,\eta_2,\eta$ carrying the ($L_e+L_\mu-L_\tau, L_s)$
charges $(-1,1),(1,1)$ and $(-2,0)$ respectively. These fields
cannot couple to $\bar{\nu}_s^c\nu_s$ but would lead to rest of
the soft terms displayed in eq.(\ref{break}). In the following, we
assume that Dirac masses $p_{e,\mu,\tau}$ are smaller than the
scale in $M_R$. This may be achieved through $R$ symmetry as in
\cite{rsym}. Here we simply make this choice which is stable
against radiative corrections.

The effective $4X4$ matrix emerging after seesaw mechanism can be
written in the form of eq.(\ref{ansatz}) with
\beq \label{msee} {\mathbf m}={1\over M_{\mu\tau}} \bmat{cc}
                 m_{e\mu}m_{\tau\tau}&m_{e\mu}(p_\tau+p_e z+p_\m z^2/y)+m_{ee}(p_e y+p_\mu z)\\
                 m_{\mu \mu}m_{\tau\tau}&m_{\mu\mu}(p_\tau+p_e z+p_\m
                 z^2/y)+m_{\mu e}(p_e y+p_\mu z)\\ \emat ~,\eeq

\beq \label{deltasee}  \Delta={1\over y M_{\mu\tau}} \bmat{cc}
                 (m_{ee}y+m_{e\mu} z)^2&(m_{ee}y+m_{e\mu} z)(m_{\mu e}y+m_{\mu\mu} z)\\
                 (m_{ee}y+m_{e\mu} z)(m_{\mu e}y+m_{\mu\mu} z)&
                 (m_{\mu e}y+m_{\mu\mu} z)^2\\ \emat ~,\eeq
\beq \label{deltapsee} \Delta'={1\over M_{\mu\tau}} \bmat{cc}
                 0&m_{\tau\tau} p_\mu\\
                 m_{\tau\tau} p_\mu&2 p_\tau p_\mu+
                 {(p_e y+p_\mu z)^2\over y}\\ \emat ~.\eeq

Here, $y=\frac{M_{\mu\tau}}{M_{ee}}$ and
$z=\frac{M_{e\tau}}{M_{ee}}$.

Comparing eq.(\ref{md}) with eq.(\ref{msee}) we find that
\beq \label{basic} \frac{a_1}{A_1}\approx \frac{a_2}{A_2}\approx
\frac{m_{e\m}}{m_{\mu\mu}} \ll 1~.\eeq
provided the elements of the Dirac neutrino mass matrix obey the
hierarchy
\beq \label{hier}
 m_{ee}\ll m_{e\mu}\sim m_{\mu e}\ll
m_{\mu\mu} ~.\eeq
This hierarchy is a natural assumption in many seesaw models. It
would follow, for example,if the  Dirac mass matrix has Fritzch
type structure. This structure also leads to

$${a_1\over A_1}-{a_2\over A_2}\sim {m_{e\m}m_{\m e}\over m_{\m\m}^2}
\frac{p_e y+p_\m z}{p_\tau+p_e z+p_\mu z^2/y}$$

This hierarchy is close to the one required on the
phenomenological grounds, see eqs.(\ref{rel1},\ref{rel2}). Thus
basic \emt symmetric ansatz is reproduced in this model under
natural assumptions.

The symmetry breaking parameters $p_{e,\mu,\tau}$ are still
arbitrary. While many choices are possible, let us identify a
particularly simple one. This corresponds to choosing $z,p_\mu$
much less than the rest. The matrices $\Delta,\Delta'$ of
eqs.(\ref{deltasee},\ref{deltapsee}) assume a simple form in the
limit $m_{ee},p_\mu,z$ tending to zero:

\beq \Delta={y\over  M_{\mu\tau}} \bmat{cc}
                 0&0\\
                 0&m_{\mu e}^2\\ \emat~~~;~~~~
\Delta'={y\over  M_{\mu\tau}} \bmat{cc}
                 0&0\\
                 0&p_e^2\\ \emat ~.\eeq

Treating non-zero elements in this matrix as perturbation we can
evaluate the mass splitting which turns out to be

\beqa \Delta_S&\approx&-2 m_1{y\over
M_{\mu\tau}}(s_\b^2p_e^2+s_\a^2 m_{\mu e}^2)
\nonumber \\
\Delta_A&\approx&-2m_2{y\over M_{\mu\tau}} (c_\b^2p_e^2+c_\a^2 m_{\mu e}^2) \nonumber \\
{\Delta_S\over \Delta_A}&\approx&
\frac{m_1}{m_2}{s_\b^2p_e^2+s_\a^2 m_{\mu e}^2\over c_\b^2
p_e^2+c_\a^2 m_{\mu e}^2} ~.\eeqa

This can be consistent with the phenomenological requirement for
proper choice of parameters.

The above exercise though illustrative in nature, shows that it is
indeed possible to integrate a singlet neutrino in seesaw picture
in a way that leads to generalize $2+2$ model and particular
ansatz discussed in section(2). Gross features of the ansatz,
particularly the required \emt symmetric matrix ${\mathbf m}$
follows under the standard seesaw assumptions. The full matrix
with a broken \emt symmetry is complex but proper symmetry
breaking can be achieved due to large number of parameters in the
model.

\section{Summary}
The three sets of experimental results namely, the observed solar
and atmospheric neutrino fluxes as well as probable oscillations
seen at LSND calls for a consistent theoretical explanation. $2+2$
schemes with four neutrinos were considered an attractive
mechanism to explain all anomalies. In these schemes, the $\nu_e$
or $\nu_{\mu}$ was exclusively assumed to convert to a sterile
state. This scenario is not supported by the results of the solar
neutrino experiment at SNO. Simple generalization which can still
explain all the experimental results assumes that sterile neutrino
flux is simultaneously but partially present in the solar as well
as atmospheric neutrino fluxes. We have shown that this
possibility follows naturally when neutrino mass matrix is \emt
symmetric. Assumption of this symmetry in fact allows us to
completely reconstruct neutrino mass matrix from the experimental
results. We have undertaken this exercise in this paper. We have
also shown that resulting structure follows naturally in two of
the conventional schemes based on Zee model and the seesaw model
for neutrino masses.\\
{\bf Note}:Before posting this paper, we noticed a preprint
(hep-ph/0110243) by K.S. Babu and R. N. Mohapatra who also
advocate the use of \emt symmetry as an explanation of the
neutrino anomalies. Model presented in this paper is different
from the models considered here.

\noindent {\bf Acknowledgements} We gratefully acknowledge
discussions with Saurabh Rindani. SG would like to thank the Theory
Division of Physical Research Laboratory for their hospitality.\\ \\

\end{document}